\documentclass{article}
\usepackage{stywhispers,amsmath,epsfig}

\usepackage{microtype}
\usepackage{graphicx}
\usepackage{subfigure}
\usepackage{booktabs} 

\usepackage{hyperref}

\usepackage{amsmath}
\usepackage{amssymb}
\usepackage{mathtools}
\usepackage{amsthm}
\usepackage{qtree}

\usepackage[capitalize,noabbrev]{cleveref}

\theoremstyle{plain}

\theoremstyle{definition}

\theoremstyle{remark}

\title{Deep Learning of Radiative Atmospheric Transfer with an Autoencoder}

\twoauthors
  {A. Abigail Basener}
	{Virginia Military Institute, \\
    Student in Applied Mathematics \\
    Lexington, VA \\
    basenerag24@mail.vmi.edu}
  {B. Bill Basener}
	{University of Virginia,\\
    School of Data Science \\
    Charlottesville, VA \\
    wb8by@virginia.edu}

\begin{document}
%
\maketitle
\begin{abstract}
As electro-optical energy from the sun propagates through the atmosphere it is affected by radiative transfer effects including absorption, emission, and scattering.  Modeling these affects is essential for scientific remote sensing measurements of the earth and atmosphere.  For example, hyperspectral imagery is a form of digital imagery collected with many, often hundreds, of wavelengths of light in pixel.  The amount of light measured at the sensor is the result of emitted sunlight, atmospheric radiative transfer, and the reflectance off the materials on the ground, all of which vary per wavelength resulting from multiple physical phenomena.  Therefore measurements of the ground spectra or atmospheric constituents requires separating these different contributions per wavelength.  In this paper, we create an autoencoder similar to denoising autoencoders treating the atmospheric affects as 'noise' and ground reflectance as truth per spectrum.  We generate hundreds of thousands of training samples by taking random samples of spectra from laboratory measurements and adding atmospheric affects using physics-based modelling via MODTRAN (http://modtran.spectral.com/modtran\_home) by varying atmospheric inputs.  This process ideally could create an autoencoder that would separate atmospheric effects and ground reflectance in hyperspectral imagery, a process called atmospheric compensation which is difficult and time-consuming requiring a combination of heuristic approximations, estimates of physical quantities, and physical modelling.  While the accuracy of our method is not as good as other methods in the field, this an important first step in applying the growing field of deep learning of physical principles to atmospheric compensation in hyperspectral imagery and remote sensing.
\end{abstract}

\begin{keywords}
Machine Learning, Deep Learning, Autoencoder, Radiative Transfer, Atmospheric Compensation
\end{keywords}

\section{Introduction}
\label{introduction}
In this paper we present a method for deep learning of atmospheric radiative transfer from a few observed spectra from a ground-based or overhead spectra.  This method could be use to measure atmospheric constituents, but we focus on using this method to convert a hyperspectral image from at-sensor radiance (the amount of light per wavelength measured at the sensor) to ground reflectance (the percent reflectance per wavelength), a process known as atmospheric compensation or atmospheric correction.  Our method uses an autoencoder~\cite{bank2020autoencoders} similar to a denoising autoencoder, treating the atmosphere as noise and ground reflectance as the image.

\begin{figure}[ht]
\vskip 0.2in
\begin{center}
\centerline{\includegraphics[width=\columnwidth]{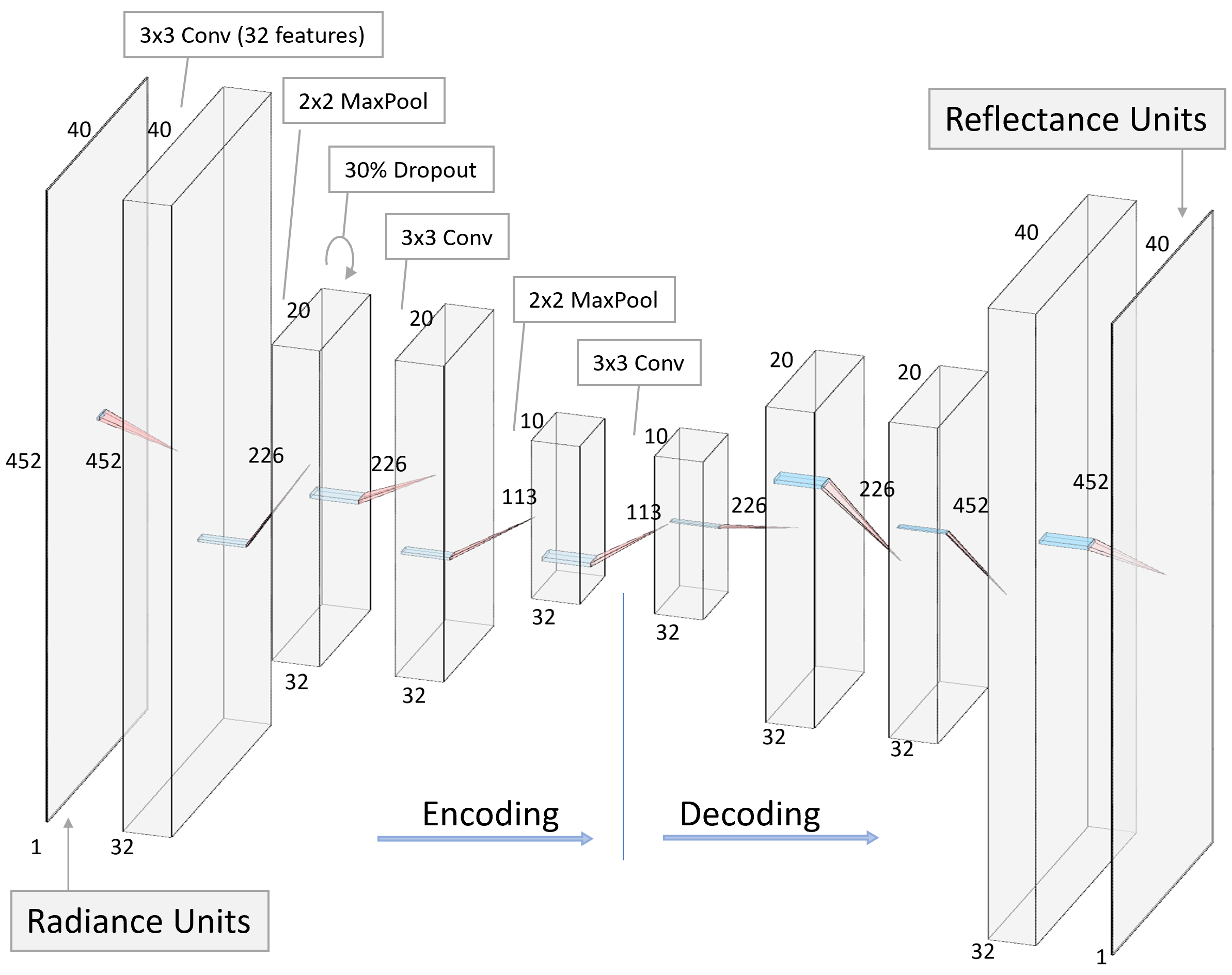}}
\caption{The autoencoder neural network used for converting from radiance to reflectance.  Details for the network architecture are included in the labels.}
\label{AuotoencoderDiagram}
\end{center}
\vskip -0.2in
\end{figure}

An autoencoder is a deep learning neural network that passes data through a series of layers that decrease in size leading to a dimensionally smaller representation of the data (encoding), and then passing the data through a symmetric series of layers back to the original data shape (decoding).  When trained well, the process learns a representation of the data in the reduced dimensional space so that the encoding stage removes noise and the decoding stage recovers the data in the original space~\cite{vincent2008extracting, vincent2010stacked, ruikai2019research, bank2020autoencoders}.  A diagram of the architecture of our autoencoder is shown in Figure~\ref{AuotoencoderDiagram}.

While our deep learning autoencoder method does not outperform a naive regression method based on the universal mean principle in QUACC~\cite{Bernstein2012}, it is an important first step in applying the growing field of deep learning of physical principles to atmospheric compensation in hyperspectral imagery and remote sensing.  We expect, based on the trajectories of other efforts in deep learning of physical phenomena, that better inclusion of physical principles in the architecture of the autoencoder would substantially improve the quality of output.  For example, the inclusion of skip connections in the ResNet Network enables layers that learn functions closer to the identity, leading to the ability to train much larger, more complex, networks~\cite{he2016deep} with a smoother loss function that has fewer local minima~\cite{li2018visualizing}. All of our data and methods are provided open access\footnote{https://www.kaggle.com/code/billbasener/autoencoder-atmospheric-compensation/notebook}.

A hyperspectral image is a digital image in which each pixel has more than the usual three color (red, green, blue) bands, but often hundreds of bands across wavelengths sufficient to get spectral information about the materials in each pixel.  We focus on hyperspectral images that have bands wavelengths from about 400nm to 2500nm - for comparison the visual colors occur around 450 (blue), 550nm (green) and 650nm (red) - and our spectra have 452 wavelengths.  For a hyperspectral images collected at these wavelengths, the measured light at the sensor is from sunlight, having passed through the atmosphere, reflected off materials, and passed again through some amount of atmosphere (which may be small for a ground-based space or significant if the sensor is on board an aircraft or satellite.)  Each band is typically 2nm to 10nm wide, and the bands are contiguous across the wavelength range.  The reflectance for a material is important because it is the result of the interaction of photons at different wavelengths and the resonant frequencies of molecular bonds (for the wavelengths above the visible range) and the interaction of photons and electrons moving between quantum states.  Specifically, important information about the constituents and bonds present in a material can be computed from reflectance spectra, for example distinguishing between different polymers, or distinguishing talcum powder from powdered sugar from dangerous white powdered substances.

\begin{figure}[ht]
\vskip 0.2in
\begin{center}
\centerline{\includegraphics[width=0.9\columnwidth]{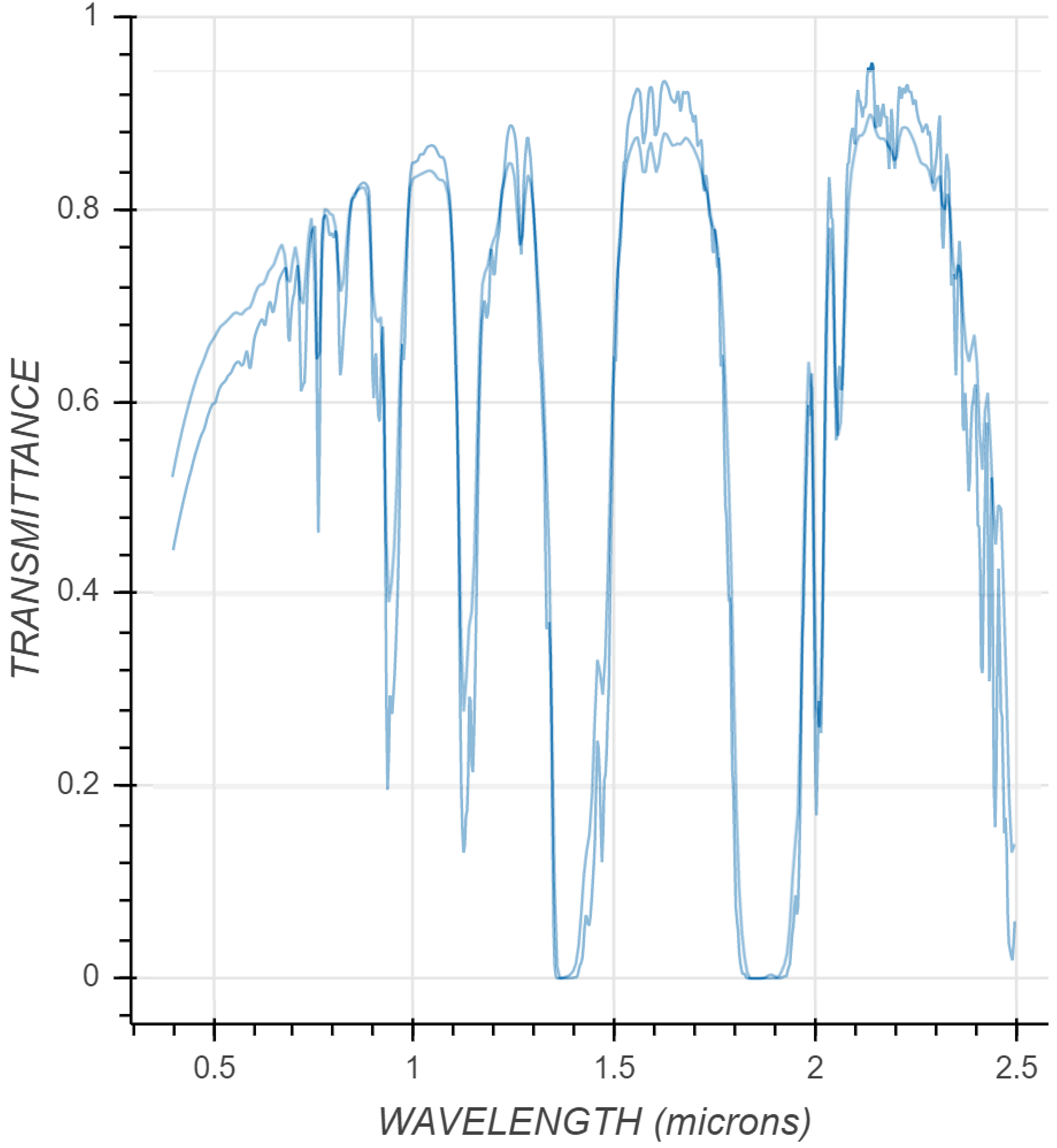}}
\caption{Two plots of transmittance for two different atmospheric models created using MODTRAN.}
\label{transmittance}
\end{center}
\vskip -0.2in
\end{figure}

The percentage of light that passes through the atmosphere (as opposed to being absorbed) is called transmittance, and varies depending on the wavelengths for each band.  A plot of transmittance for two different atmospheric models across our wavelength range with a spectral resolution (band width) of 5nm is shown in Figure~\ref{transmittance}.  Part of our goal is to determine this transmittance amount from spectra on the ground measured from a sensor even when the reflectance of the ground material is unknown.

\begin{figure}[ht]
\vskip 0.2in
\begin{center}
\centerline{\includegraphics[width=0.9\columnwidth]{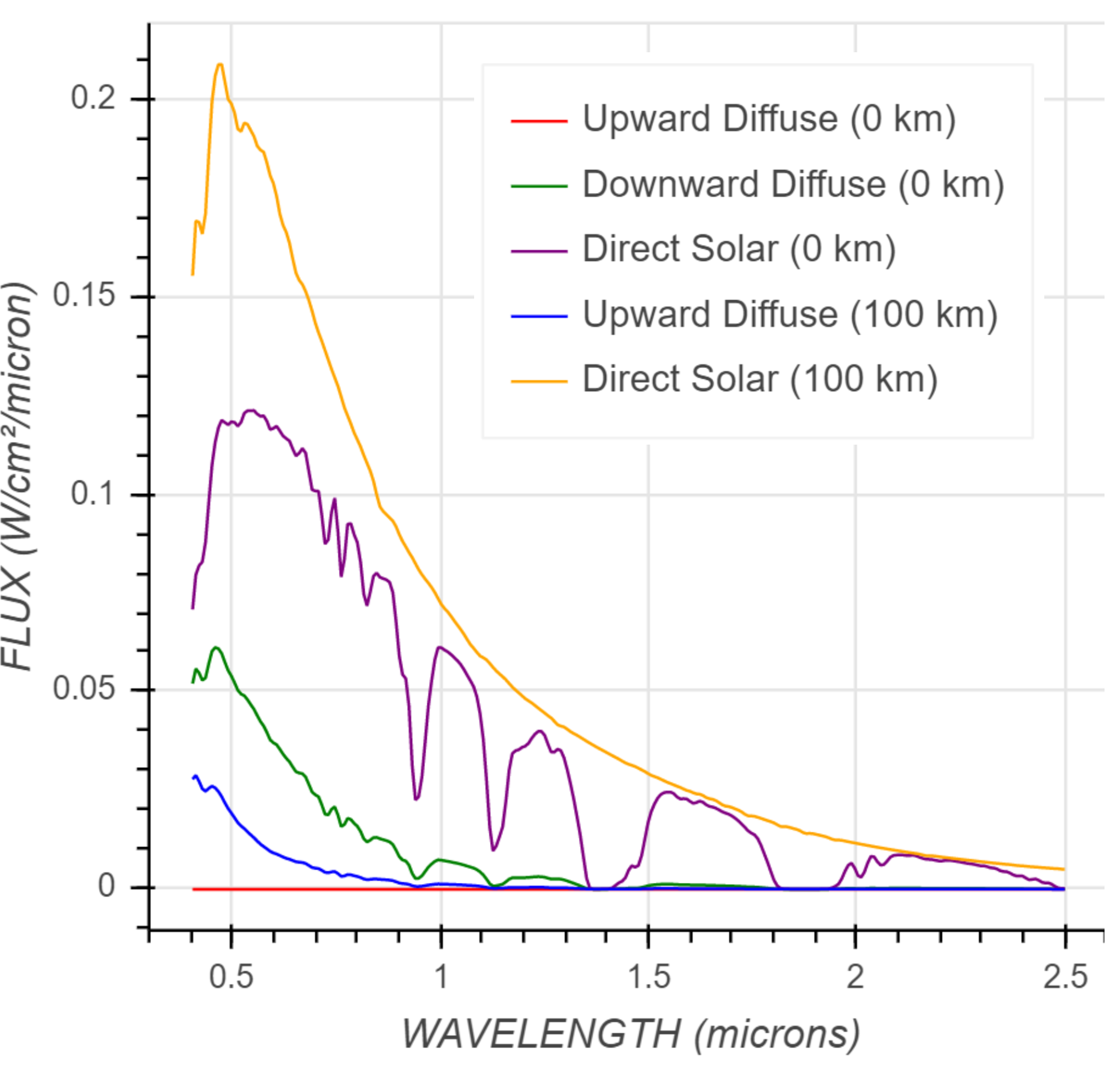}}
\caption{Two plots of transmittance for two different atmospheric models created using MODTRAN.}
\label{flux}
\end{center}
\vskip -0.2in
\end{figure}

The plot in Figure~\ref{flux} shows more components involved in the radiative transfer model  The Direct Solar (100km) gives the amount of sunlight measured at the top of the atmosphere.  The main shape of this curve is from the blackbody radiation given the temperature of the sun.  The Direct Solar (0km) is the amount of sunlight reaching the ground, which is the Direct Solar (100km) times the transmittance shown in Figure\ref{transmittance}.  The Downward Diffuse (0km) is the amount of light per wavelength that reaches the ground after scattering in the atmosphere; this is the indirect illumination on an object that is in a shadow from the sun.  The upward Diffuse is the amount of upward light, which at 100mk (top of atmosphere) is from atmospheric scattering and at 0km is from the blackbody radiation of the ground (which is insignificant given the wavelength range and ground temperature assumed in this model).

The at-sensor radiance is the Upward Diffuse (at the elevation of the sensor) plus the total illumination (Direct Solar + Downward Diffuse) times the percent reflectance per band of the material on the ground per wavelength $_\lambda$,
\[
UD_\lambda + (DS_\lambda + DD_\lambda )Ref_\lambda = Rad_\lambda.
\]
There are also nonlinear effects as well.  Light that has passed through a leaf and reflected off the ground would have the leaf transmition times ground reflectance in place of $Ref_\lambda$ in this equation.  In the lower wavelengths, especially blue and below, photons will take multiple bounces/scattering (collectively, 'haze' in the image) in the atmosphere (which you can observe in the higher values for Upward/Downward Diffuse in Figure~\ref{flux}), but the amount of haze is highly dependent on constituents and length of pass through the atmosphere; moreover has causes photons reflecting off material at one location to scatter in the atmoshpere and enter the sensor at locations/directions for other pixels, causing in "Upward Diffues" that varies across the image comprised of a nonlinear mixture of nearby ground material spectra and atmospherics, rather than the ideal Upward Diffuse from the atmosphere alone shown in figure~\ref{flux}.

All of these values change with respect to atmospheric constituents, water vapor, CO2, Ozone, CH4, aerosols, sun angle, fraction of sun and sky visible to each pixel (shadow from objects, terrain, clouds, etc.), sensor angle, angle and roughness of the ground material, and other factors.  The MODTRAN software (http://modtran.spectral.com/modtran\_home) can simulate these effects if they are known, and provide a modeled at-sensor radiance for reflectance spectra, or a ground reflectance spectrum for a measured at sensor radiance.

The purpose of hyperspectral imaging is to perform spectroscopy writ large; that is, to be able to determine the materials in each pixel from the reflectance spectra for those pixels.  As such, good atmospheric compensation is an essential step.  The most accurate methods are usually either based on physics based modelling with MODTRAN (MODerate resolution atmospheric TRANsmission, http://modtran.spectral.com/modtran\_home) such as FLAASH~\cite{Perkins2012} or using materials of known reflectance in the image, for example the empirical line method~\cite{mamaghani2018initial,gao2009atmospheric}, in which case it is usually preferable to have materials that are spectrally flat, for example a set of five panels which are 5\%, 30\% 50\% 80\% and 95\% reflectance across all wavelengths, which can be used to estimate a best fit regression line per wavelength to convert from radiance units to reflectance.  However, all of these methods have some approximations and attempt to measure physical parameters.  For example, a good ELM will estimate the Upward diffuse as the intercept and the Direct Solar and Downward Diffuse as the slope, but assume these are consistent values for every pixel in the image.  FLAASH attempts to estimate the physical parameters from the image, even estimating water vapor content per pixel, and use a physics based model to compute the grouped reflectance from each at-sensor radiance measurement.

A heuristic and approximate but surprisingly effective method for atmospheric correction is to make the assumption that the mean spectrum of a significantly large library of reflectance spectra will be consistent, and use this assumption to compute a single gain and offset that is applied per wavelength across the image.  A method based on this assumption is QUACC (QUick Atmospheric Correction Code), in which a sample of 50 different spectra (called endmembers) are selected from the radiance image, usually iteratively so each new endmember is optimally different than the previous ones, and the mean of these 50 endmembers is determined.  The least measured radiance value across the image per wavelength is assumed to be the upwelling radiance, and is thus subtracted from the image.  Then the ratio between the ideal mean and mean of the endmembers (after subtracting estimated upwelling radiance) is computed and used as a 'gain' and is multiplied by every upwelling-subtracted radiance value to convert to reflectance.  This is often implemented with additional heuristic improvements such as removing spectra of mud or vegetation from the spectra, is provided in QUACC~\cite{Bernstein2012}.  There are other semi-heuristic methods for example SHARC~\cite{Katkovsky2018}.  The QUAAC method is faster than physics-based methods are quires no manual input.  In tests, is provides reflectance spectra that are $\pm 15\%$ in comparison to FLAASH generated reflectance spectra, and perhaps more importantly QUACC tends to generate reflectance spectra that retain the features of the true spectra, which is the most important factor for spectroscopy.  A small to moderate variation in the magnitude of a spectrum is often not important.  Materials are identified more from variation in reflectance at different bands which indicates which bands are absorbing vs. reflecting, resulting from the constituents and molecular bonds present.  The total reflectance (i.e. total albedo) can vary with illumination amount, angle to sun, and surface roughness, inconsistent calibration, none of which are dependent on the molecular bonds and constituents of the material.  The primary questions in atmospheric compensation methods is whether they retain the features of the true reflectance and whether they avoid creating new features not present in the material.

\begin{figure}[ht]
\begin{center}
\centerline{\includegraphics[width=0.8\columnwidth]{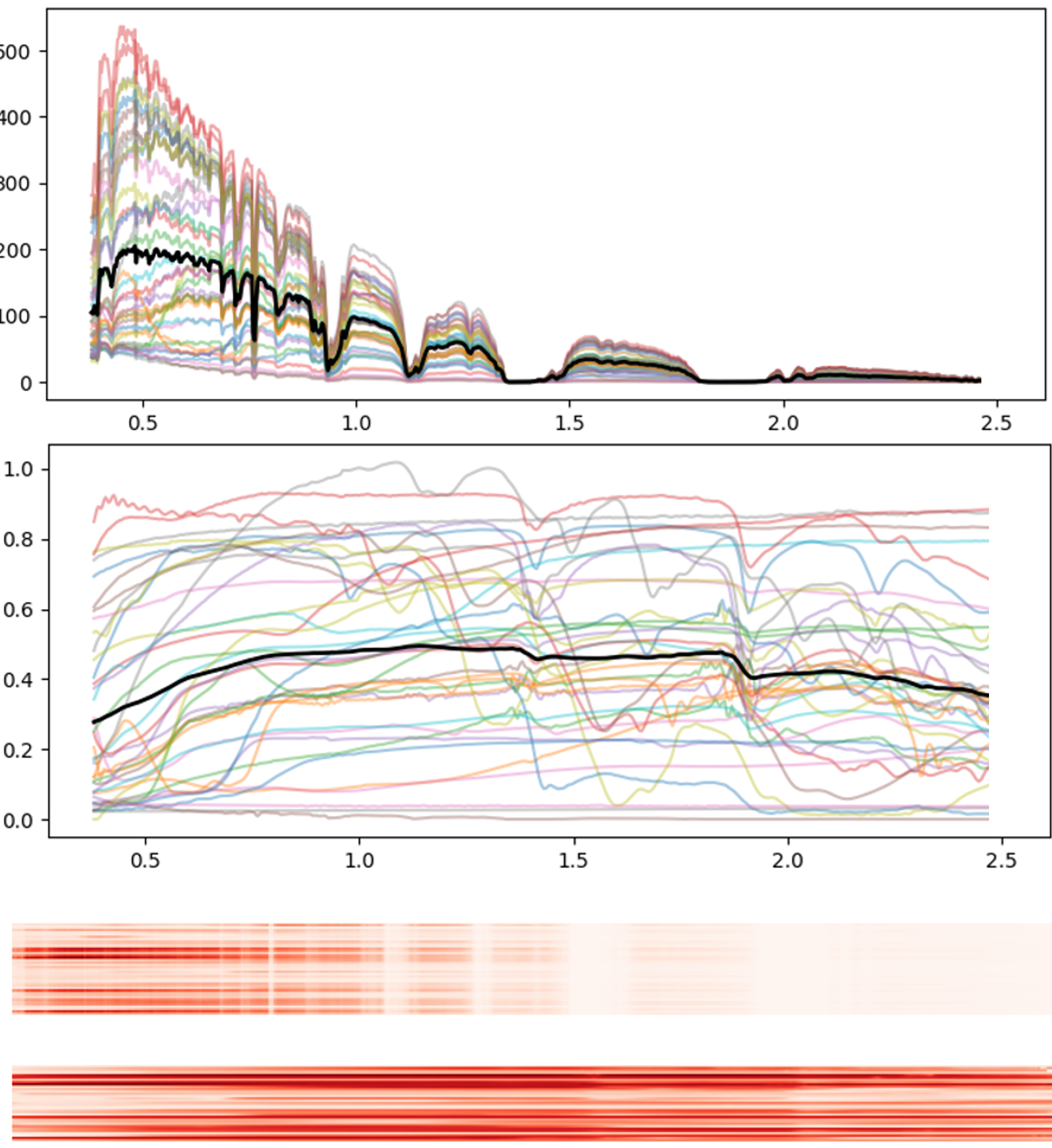}}
\caption{At-sensor radiance for 39 spectra and the mean plotted in black (top plot) along with the true reflectance for these materials and the mean of their reflectance in black (bottom plot).  Below are the data arrays for these plots (40 spectra by 452 bands each) are shown below as images.}
\label{SampleRadAndRef}
\end{center}
\vskip -0.2in
\end{figure}

For this paper, we started with a set of about 5,000 reflectance spectra of known materials, each of which passed some basic quality check.  We then took random samples of size 39, randomly selected a set of parameters for MODTAN (solar zenith angle from 0-85 in increments of 5, random selection from the 6 possible atmospheric models, random selection from the 12 possible aerosol models) and created an associated set of 39 radiance spectra.  For each set, we also computed the mean spectrum and added this as a $40^\textrm(th)$ spectra.  This creates arrays that have even dimension (number of spectra and number of bands) which are preferable for some Deep Learning methods.  So our input data is a set of 40 radiance spectra each with 452 wavelengths, and our output data is a set of 40 reflectance spectra.  In Figure~\ref{SampleRadAndRef} we show a plot of the 40 radiance spectra (top) and the associated reflectance spectra (bottom).  We created a set of 10,000 such samples of 40 spectra.

\section{Baseline Regression Method and Results}
We create a baseline regression method for atmospheric correction based on the assumption that the mean spectrum of a significantly large library of reflectance spectra will be consistent, called the reference spectrum.  The is the same assumption used in the QUACC~\cite{Bernstein2012} method, although we are using the synthetically generated data which has different distributions for the types of spectra present.  Our baseline is not an approximation to QUACC, but a baseline for comparison to the Deep Learning methods that is simple and based on an accepted heuristic approximation.

\begin{figure}[ht]
\vskip 0.2in
\begin{center}
\centerline{\includegraphics[width=\columnwidth]{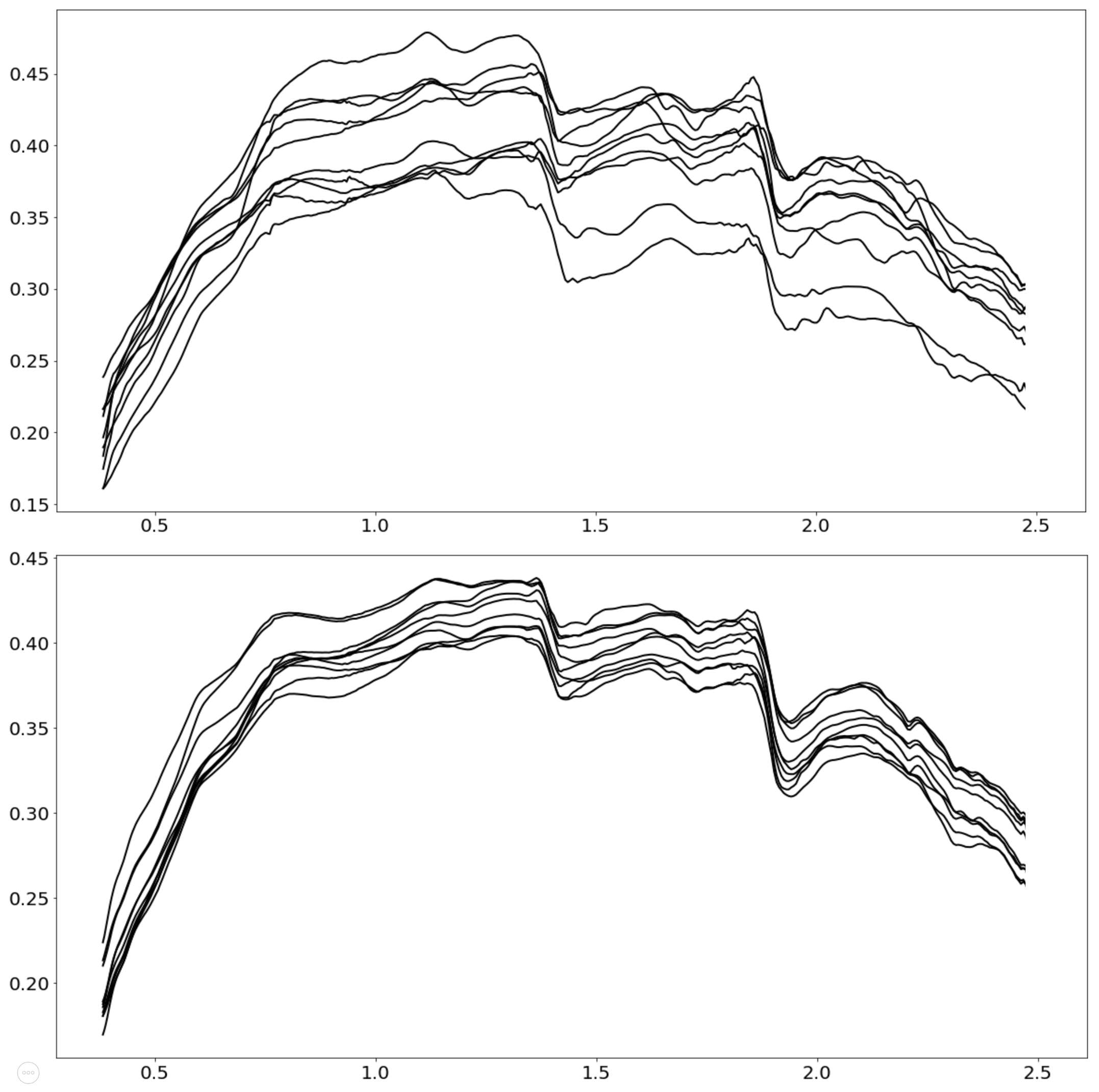}}
\caption{Ten mean spectra, each of which is the mean from a random selection of 40 reflectance spectra (top) and 120 spectra (bottom).}
\label{meanReflectanceSpectra}
\end{center}
\vskip -0.2in
\end{figure}

The mean spectra for a each of a random selection of ten of our sets of 40 spectra are shown plotted in Figure~\ref{meanReflectanceSpectra}.  This assumption is very approximately true, but the approximation seems more consistent using 120 per sample, shown below in this figure.
A physics-based support for the general shape of the reference reflectance spectrum is provided in~\cite{Bernstein2012} as follows:
\begin{quote}
The general shape of the reference reflectance spectrum has a simple physical origin. The decrease toward the longwavelength edge arises because the molecular
constituents of materials have relatively strong NIR vibrational absorption features that increase in strength with increasing wavelength. The decrease toward the short-wavelength edge arises because the molecular constituents have strong electronic absorption features that increase in strength with decreasing wavelength. Although we normalize the peak of this curve to unity for reasons discussed later, it is important to note that the peak average reflectance is $\sim 0.4$...
\end{quote}

\begin{figure}[ht]
\vskip 0.2in
\begin{center}
\centerline{\includegraphics[width=\columnwidth]{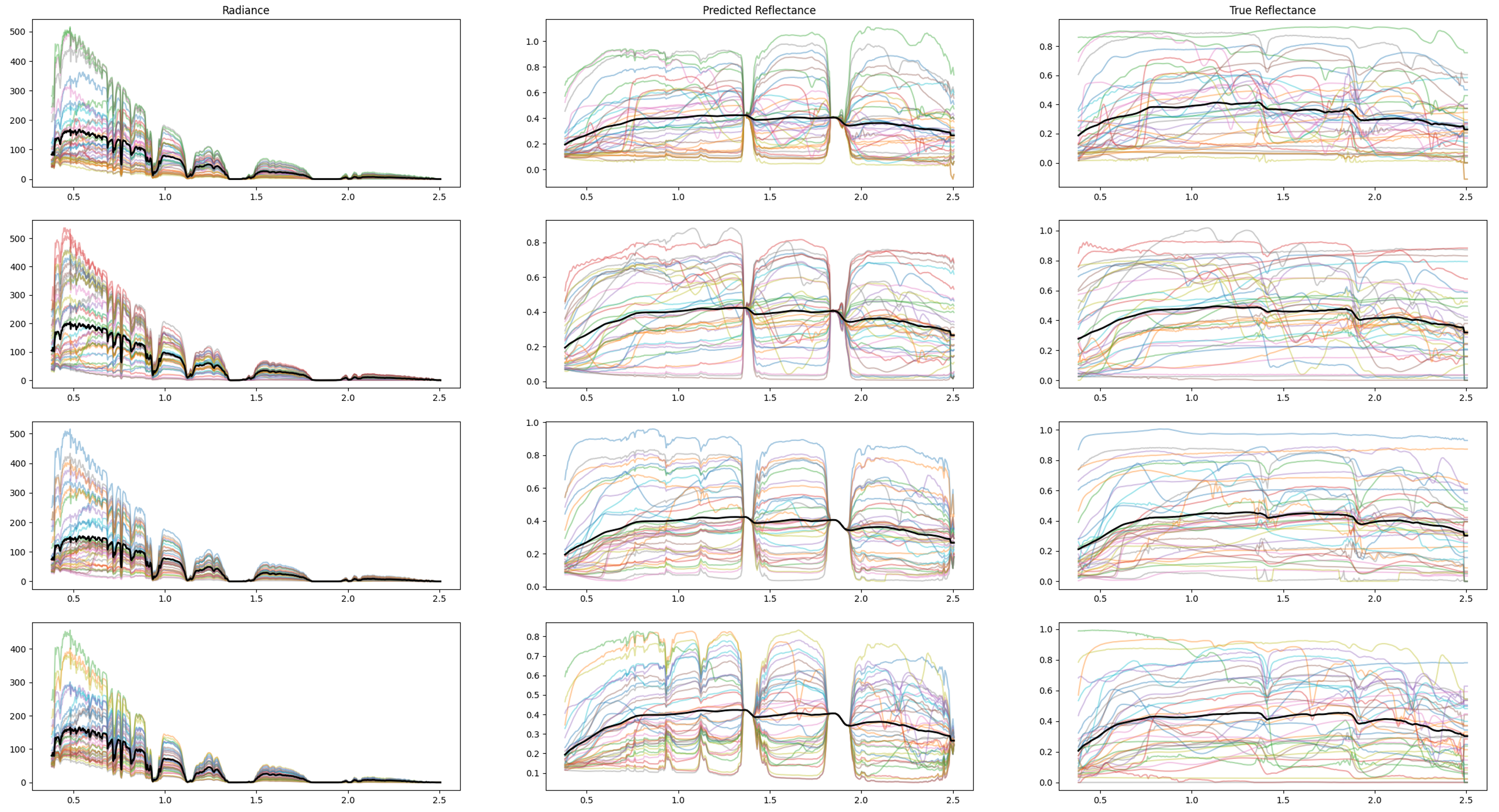}}
\caption{Three samples of spectra in modeled radiance units (first column), converted to reflectance using the heuristic gain from baseline method (center column), and true reflectance (last column).  The water absorption regions (just below 1.5 and just below 2.0) are not errors but locations where the water absorption in the astmosphere effectively blocks all light.}
\label{test_results_regression_method}
\end{center}
\vskip -0.2in
\end{figure}

\begin{figure}[ht]
\vskip 0.2in
\begin{center}
\centerline{\includegraphics[width=\columnwidth]{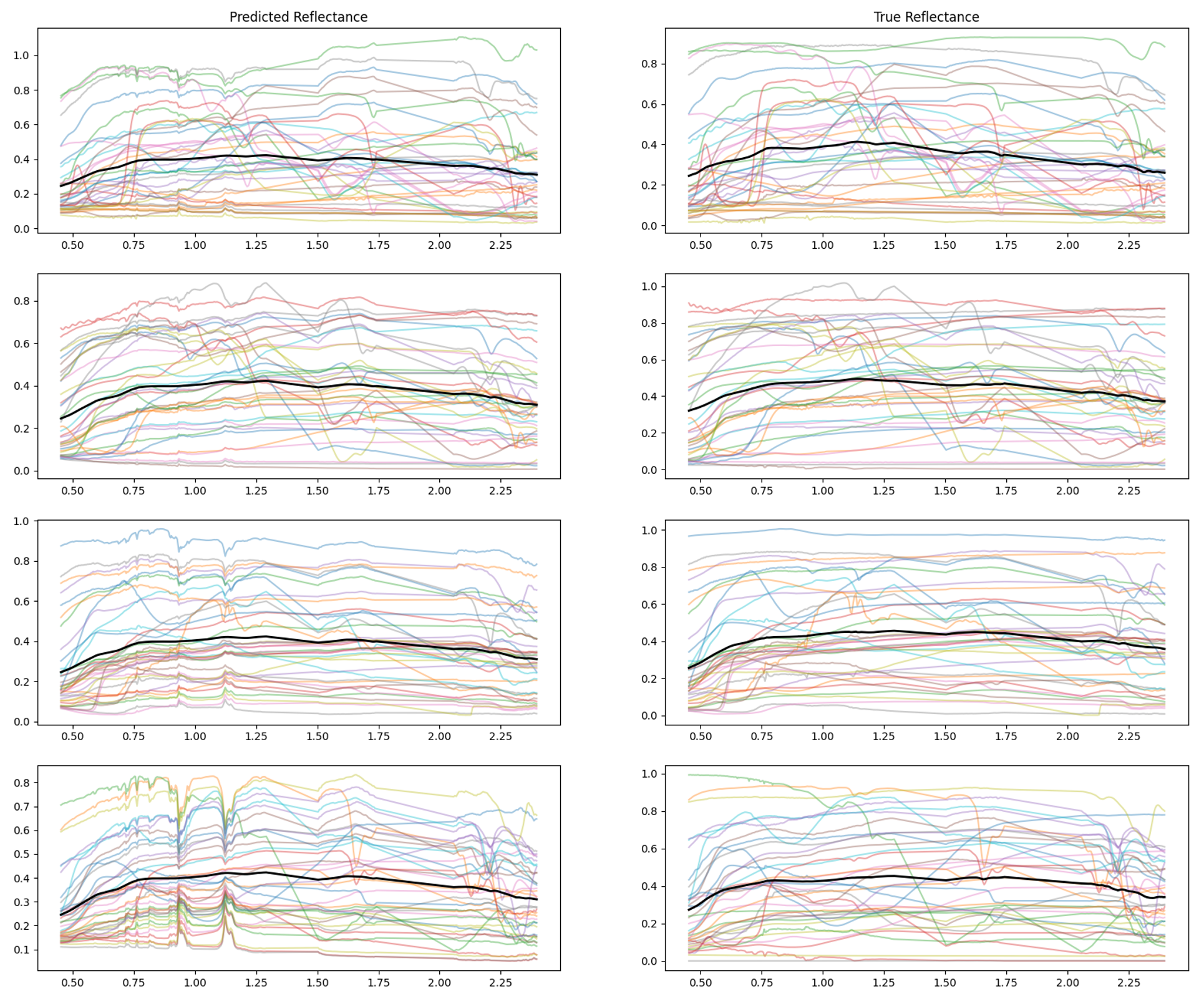}}
\caption{Predicted reflectance using the heuristic gain from baseline method (left-hand column), and true reflectance (right-hand column).  The water absorption regions have been removed (replaced with a straight line in plots) to aid in visual comparison of features.  Visual inspection shows that many of the features in the true reflectance are present in the predicted reflectance.}
\label{test_results_regression_method_bbr}
\end{center}
\vskip -0.2in
\end{figure}

From the assumption that a sample of spectra will have a fixed mean reflectance $m$, we can simply take this mean divided the mean measured radiance spectra to obtain a 'gain' multiplication factor for each band.  An approximate atmospheric compensation can be done by multiplying this gain times each measured radiance value, per band.  The output from the baseline method is shown in Figure~\ref{test_results_regression_method}.  Some of the features in the true reflectance spectra column (third, right-hand, column)can be observed in the spectra predicted using the baseline method with the standard mean reflectance assumption (center column).  There are regions in the predicted column where the spectra have values near zero - these are not errors, but regions where the radiance has near-zero values.  Effectively, the water in the atmosphere has near-zero transmittance in these wavelength regions, and there is insufficient signal for prediction.  These regions are removed from all hyperspectral images created with solar illumination.  Wavelength regions around 0.9 and 1.2 microns are also often removed because some atmospheres will also have very low transmittance in these regions, and these regions can be observed particularly strong in the 4th row.  Figure~\ref{test_results_regression_method_bbr} shows these same spectra but with the water bands removed (replaced with a straight line in the plots) to aid in visually comparing spectra.

To measure the effectiveness of this method we randomly selected 500 sets of 39 spectra and computed the correlation between each true reflectance and predicted reflectance (after removing bands in $0-0.45$,$1.3-1.5$,$1.75-2.05$, and $2.4-3$ micron regions) and the mean correlation of these 19,500 spectra was 0.87 with a standard deviation of 0.29.  Of these spectra, $69\%$ of the predicted spectra were within $15\%$ of the true spectrum.  These results are not great, and not sufficient without addition modifications like those in QUACC, but this is sufficiently promising that a deep learning method may be viable.

\section{Autoencoder Method and Results}
Because the simple baseline regression method works reasonably, it can be expected with a proper architecture and proper training, a deep learning method could improve on accuracy.  Perhaps the model could learn the baseline regression plus adjustments based on common types of variation in the atmosphere and illumination.  Because the in put and output have the same shape, we decided to try an autencoder network treating the atmosphere as noise (see~\cite{bank2020autoencoders}).

The input into our autencoder is the $40\times452$ array of 40 spectra in radiance units, each of which has 452 bands.  The output is an array of the same shape, number of spectra, and bands, but in reflectance units.  The architecture for the network is shown in Figure~\ref{AuotoencoderDiagram}, constructed in Keras and trained for 50 epochs in batches of 256 each with an adam optimizer and binary\_crossentropy loss.  All activation functions are ReLU except the final decoder layer which is sigmoid, in which case the $0-1$ output of the sigmoid neurons match the $0-1$ range of values for reflectance.

\begin{figure}[ht]
\vskip 0.2in
\begin{center}
\centerline{\includegraphics[width=\columnwidth]{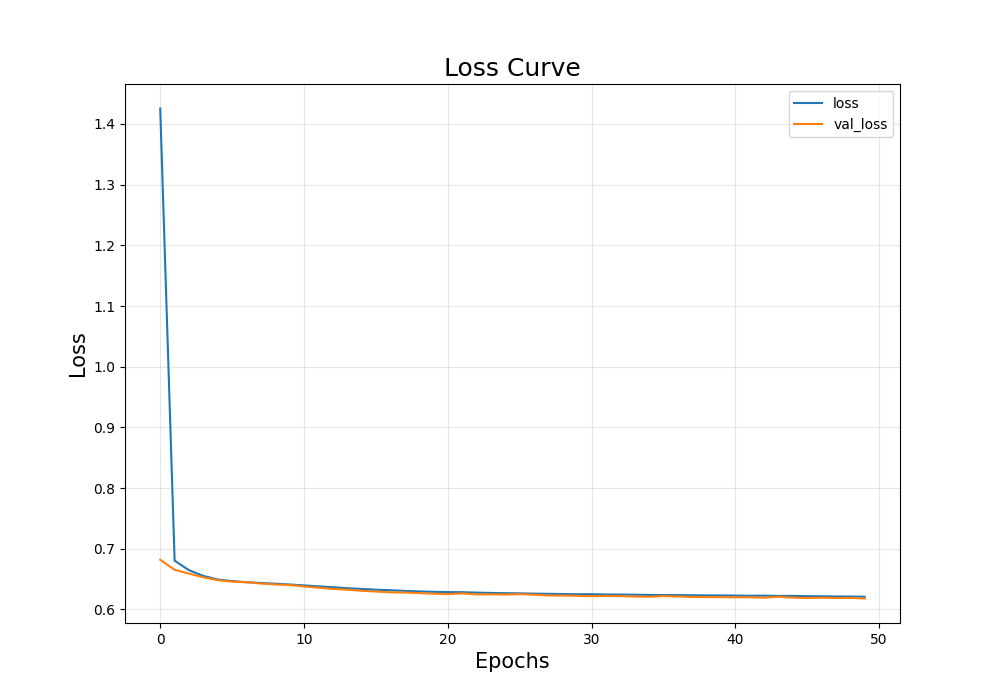}}
\caption{The loss curve from training the autoencoder.}
\label{Loss_curve}
\end{center}
\vskip -0.2in
\end{figure}

\begin{figure}[ht]
\vskip 0.2in
\begin{center}
\centerline{\includegraphics[width=\columnwidth]{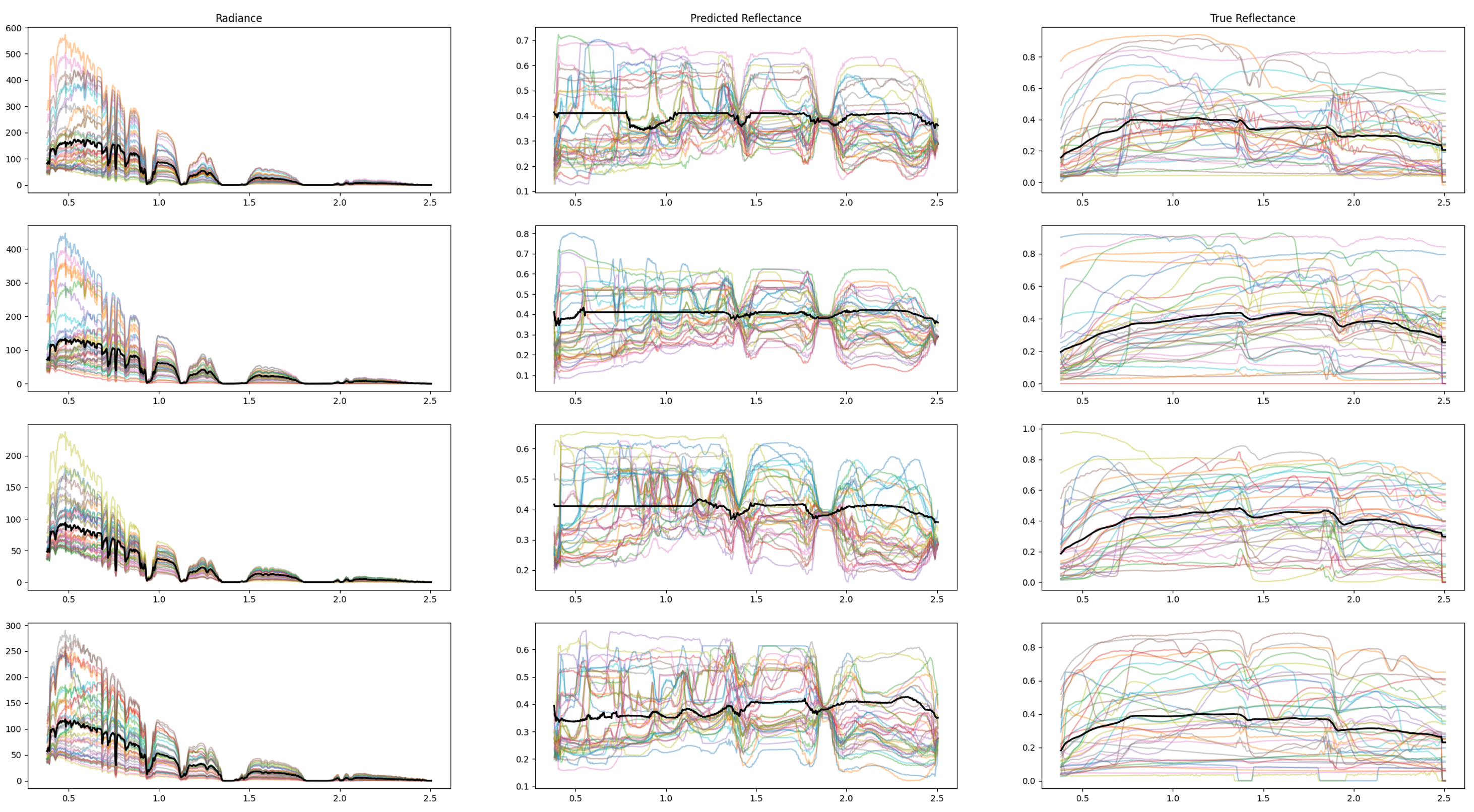}}
\caption{Three samples of spectra in modeled radiance units (first column), converted to reflectance using the autoencoder (center column), and true reflectance (last column).  The water absorption regions (just below 1.5 and just below 2.0) are not errors but locations where the water absorption in the astmosphere effectively blocks all light.}
\label{test_results}
\end{center}
\vskip -0.2in
\end{figure}

\begin{figure}[ht]
\vskip 0.2in
\begin{center}
\centerline{\includegraphics[width=\columnwidth]{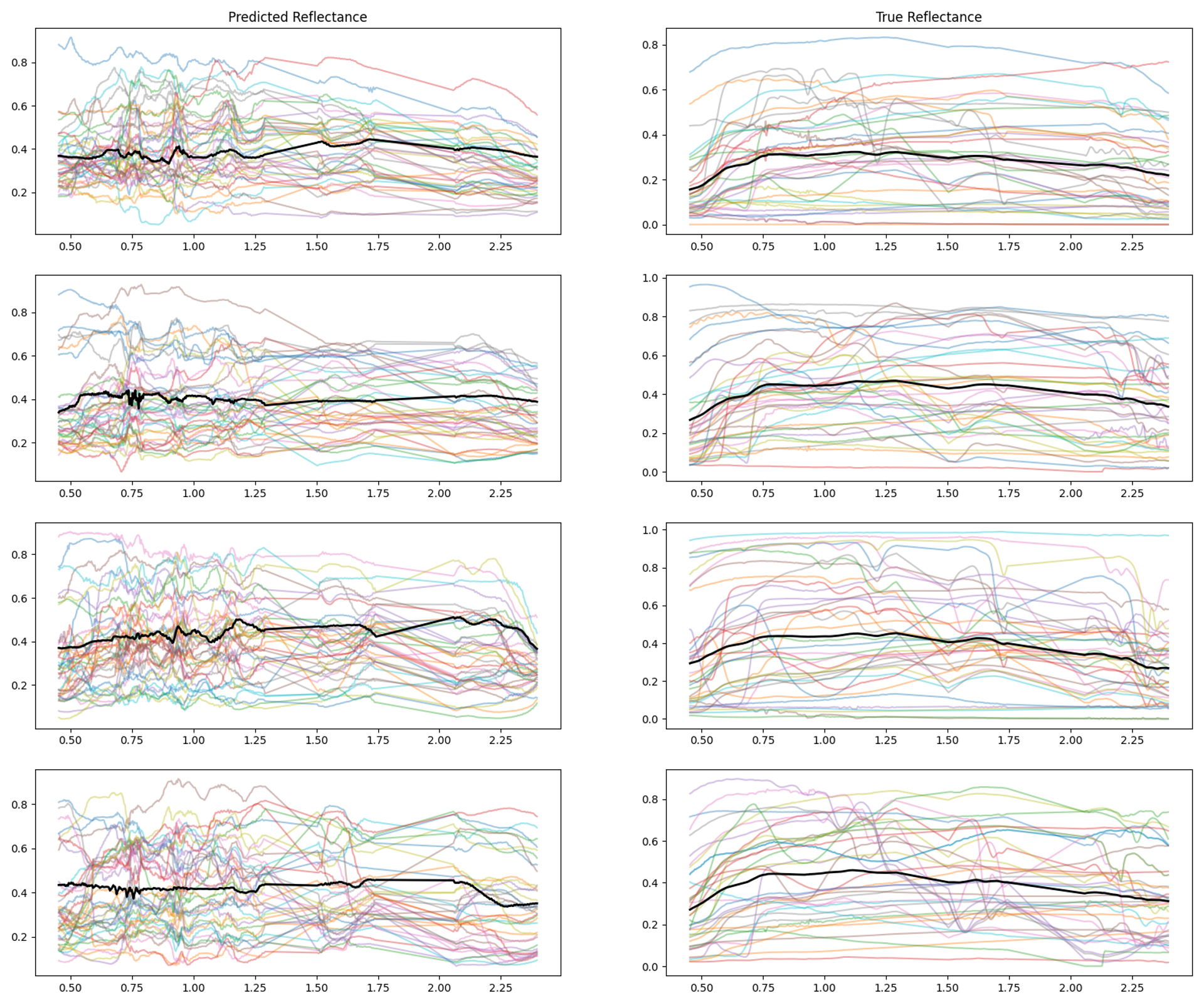}}
\caption{Predicted reflectance using the autoencoder (left-hand column), and true reflectance (right-hand column).  The water absorption regions have been removed (replaced with a straight line in plots) to aid in visual comparison of features.  Observe that, especially in the 0.45 to 1.25 micon region, the autoencoder is introducing features that are not present in the true reflectance spectra.}
\label{test_results 2_bbr}
\end{center}
\vskip -0.2in
\end{figure}

The loss curve from optimizing the network is shown in Figure~\ref{Loss_curve} (using $33\%$ of the data for validation).  Example output showing original radiance, predicted reflectance, and true reflectance are shown in Figure~\ref{test_results}.  The output predicted reflectance of the autoencoder along with true reflectance with the standard water bands removed are shown in Figure~\ref{test_results 2_bbr}.

To measure the effectiveness of the autoencoder comparable to our baseline method, as before we randomly selected 500 sets of 39 spectra from the validation set and computed the correlation between each true reflectance and predicted reflectance (after removing bands in $0-0.45$,$1.3-1.5$,$1.75-2.05$, and $2.4-3$ micron regions) and the mean correlation of these 19,500 spectra was 0.52 with a standard deviation of 0.44.  Of these spectra, $34\%$ of the predicted spectra were within $15\%$ of the true spectrum.  These results are significantly worse than the baseline method.

\section{Conclusions}
It is clear that our autoencoder performed poorly in comparison to the baseline method, summarized in Table~\ref{results_table}.
\begin{table}[t]
\caption{Evaluation metrics for the baseline regression and autoencoder methods.}
\label{results_table}
\vskip 0.15in
\begin{center}
\begin{small}
\begin{sc}
\begin{tabular}{lcccr}
\toprule
Metric & Regression   & Autoencoder   \\
  &   Result &   Result \\
\midrule
Mean Corr                   & 0.87      & 0.52   \\
Std Corr                    & 0.29      & 0.44   \\
percentile in $\pm 15\%$    & $69\%$    & $34\%$ \\
\bottomrule
\end{tabular}
\end{sc}
\end{small}
\end{center}
\vskip -0.1in
\end{table}

It seems likely that the architecture of CNNs is making the network unable to use the wavelength information.  Specifically, the architecture we used folllows a standard framework for images.  In images, a combination of pixels forming a nose/ear/wheel/etc. is meaningful at any location in the image, and CNNs are able to leverage this information.  But in spectroscopy, the meaning of a combination of values is dependent on the location of the feature; although a network might learn the shape of features in general and use both the shape and location in final layers.  We tried a significant number of modifications to the architecture, including replacing some or all Conv layers with fully connected dense layers, modifying the amount of dropout, using different optimization methods and loss functions (for example, rmsprop in place of adam and/or mean squared error in place of binary crossentropy).  We also tried building the network to use just the mean spectrum (to learn an approximation to the baseline regression concept), or building convolution layers with different rectangular windows.  However, none of these methods provided benefit above the autencoder provided in this section in any manner worth reporting.  It does seem the architecture should learn a function for removing atmosphere in a way that the same function is applied to every row (spectrum) in the data array.

Our approach was to use an autoencoder and treat the atmosphere as noise since this is a well-developed method of application.  But perhaps since the baseline regression method is a reasonable approximation, perhaps a regression neural network would be better, or some combination that leverages the known heuristics and physics together with a deep learning approach.  We believe the physical explanation of the data and reasonable effectiveness of the baseline regression method strongly suggest that a Deep Learning approach to atmospheric compensation, perhaps incorporated with some heuristics and physics, has the likelihood of being completely automated and highly effective, with the potential to outperform currently available methods such as QUACC and FLAASH.
\bibliography{my_refs4}
\bibliographystyle{IEEEbib}

\end{document}